\documentclass[11pt]{revtex4}

\usepackage{graphicx}
\usepackage{dcolumn}
\usepackage{bm}

\def\al{\alpha}

\def\be{\begin{equation}}
\def\ee{\end{equation}}
\def\bea{\begin{eqnarray}}
\def\eea{\end{eqnarray}}
\def\la{\label}

\def\bsea{\begin{subeqnarray}}
\def\esea{\end{subeqnarray}}

\def\tr{\mbox{tr}}

\begin{document}


\title{ Out of Equilibrium Dynamics of the Toy Model with Mode Coupling and Trivial Hamiltonian
\footnote{This article is dedicated to Bob Dorfman on his 65th birthday.
His persistent contributions to the kinetic theory and to the bases of 
nonequilibrium statistical physics over more than three decades taught us a great deal
on the subjects and on the style of doing science.} }

\author{Kyozi Kawasaki}
\altaffiliation{CNLS, Los Alamos National Laboratory, Los Alamos, NM 87545, USA\\
 Phone:505-667-9469, fax:505-665-2659, e-mail:kawasaki@cnls.lanl.gov \\
Permanent address: 4-37-9 Takamidai, Higashi-ku, Fukuoka 811-0215, Japan\\
Phone and fax:+81-92-607-4022, e-mail:kawasaki@athena.ap.kyushu-u.ac.jp}
\author{Bongsoo Kim}
\altaffiliation{Department of Physics, Changwon National University, Changwon 641-773, Korea \\
Phone:+82-55-279-7417, fax:+82-55-267-0264, e-mail:bskim@sarim.changwon.ac.kr}

\affiliation{%
}%


\date{\today}

\begin{abstract}
{\bf Abstract} \\
We extend our previous analysis of the toy model that mimics
the mode coupling theory of supercooled liquids and glass 
transitions to the out of equilibrium dynamics. 
We derive a self-consistendt set of equations for correlation and response functions. \\

\end{abstract}

\maketitle

\section{Introduction}

Recently we have introduced a mean field toy model that mimics the
mode coupling theory(MCT) of supercooled liquids and glass transitions 
with trivial Hamiltonian \cite{kkbk,kik01,merida}. Analyses were limited to the
equilibrium dynamics. An important feature of the model is that the
strength of ``hopping processes'' \cite{goe} that destroys the non-ergodic state of
the ideal MCT \cite{leuth,bgs} can be tuned so that nonergodic state is still allowed in
some region of the model parameter space. This implies that the
so-called hopping processes do not seem to be the same as thermally
activated processes.
 
In order to obtain further insights into the nature of MCT we consider the
out of equilibrium dynamics of the model.
In connection to this, recently the out of equilibrium dynamics of the
 mean-field-type spin glass models  and other related glassy models 
was considered \cite{trieste}. In particular, the out of equilbrium dynamics of 
the spherical $p$-spin model \cite{crisanti}, represented by the closed set of 
equations for the off-equilibrium two time correlation function $C(t,t_w)$ and the
response function $G(t,t_w)$, was analytically solved in the long time regime \cite{ck}.
The analytic solution has revealed  interesting features of  out of equilibrium 
dynamics of the model. The system exhibits a strong waiting time dependence
in the relaxaion of both $C$ and $G$, i.e., aging behavior at low temperatures.
Moreover, the fluctuation-dissipation theorem (FDT), i.e., 
the relationship bewteen $C$ and $G$ in equilbrium, is modified in an interesting way.
Similar FDT violation have been observed in
the off-equilibrium dynamics of supercooled liquids in computer simulations
 \cite{kobarrat,parisi,leo} and an experiment \cite{israel}.

We note that the all the out of equilbrium glassy features in the $p$-spin and related 
models are driven, as in the equilibrium dynamics, by the dissipative nonliearity 
in the equation of motion which comes from the nonlinear Hamiltonian.
In the present toy model, as we see below, there is no dissipative nonlinearity 
since the Hamiltonian is trivial, i.e., gaussian without disorder.
Instead the equation of motion involves the non-dissipative, i.e., reversible
mode coupling nonliearities which drives the slowing down in the relaxation and the
dynamic transition in the equilibrium dynamics.
Our model possesses the reversible nonlinearities since we had a fluid in mind 
in constructing the model. 
Here we aim to see the out of equilibrium dynamics of the model 
driven by these reversible nonlinearities. 

As a first step in this direction we derive below the
self-consistent closed set of equations for five correlation functions
and five response functions. The method used is standard: the
generating functional method in which two fictitious external fields
are introduced for each dynamical variables entering the
model \cite{bckm}. But in view of more complications involving
correlation and response functions of these variables we will sketch
a derivation. We indicate possibility of reducing these complicated
equations to the set of two correlation functions and two response functions.
\section{TOY MODEL}
Here we consider the toy model with $M$-component velocity-like $b$
variables and $N$-component density-like  $a$ variables, $M$ being smaller
than $N$.  We have shown that in the limit of $M, N \rightarrow \infty$
with $\delta^* \equiv M/N$ finite, the parameter $\delta^*$  becomes a
measure of hopping \cite{goe}. As a special case, if we take $\delta^*=0$, then we
obtain the zero-hopping model for the variables $a$, that is, the model is trivially non-ergodic. 
For $\delta^*=1$ the hopping fully contributes and the system is always ergodic.
For intermediate values of $\delta^*$
we expect an ergodic to nonergodic transition at some value of $T=T(\delta^*)$.

In this paper, after introducing the model we consider off-equilibrium dynamics that 
eventually involves 5 correlation functions and 5 response functions.

Our toy model is described by the variables
$a_j$ with $j=1,2, \cdots, N$, and $b_{\alpha}$ with $\alpha=1,2, \cdots,
M$ which are sometimes abbreviated as $\hat x$ which spans the phase space of
the model.  The model is described by the following
Langevin equation
\bea
&{}&\dot{a}_i= K_{i\alpha}b_{\alpha}+\frac{\omega}{\sqrt{N}}
J_{ij\alpha}  a_j b_{\alpha} \nonumber \\
&{}&\dot{b}_{\alpha} =-\gamma b_{\alpha}-\omega^2 K_{j\alpha}a_j-
\frac{\omega}{\sqrt{N}}J_{ij\alpha} (\omega^2a_ia_j- T\delta_{ij})+f_{\alpha}  \nonumber \\
&{}& <f_{\alpha}(t)f_{\beta}(t')> = 2\gamma T\delta_{\alpha \beta} \delta (t-t') 
\la{eqn:s6}
\eea
where the $f$'s is the thermal noise with zero mean and the angular bracket is the thermal
average over such noise, and the usual summation convention for repeated indices are used.
Here and after we will use Roman indices for the component of $a$
and Greek for that of $b$.
Here $\gamma$ gives a decay rate of the variable $b_{\alpha}$ and $\omega$ is
 seen to give a measure of the frequency of the oscillation of
the variable $a_j$.

 For later purpose, we require that the matrix
$K_{i\alpha}$ satisfies $K_{i\alpha} K_{i\beta}=\delta_{\alpha
\beta}$. It is then  easy to show that the equilibrium stationary phase
space distribution of the Fokker-Planck equation corresponding to  (\ref{eqn:s6}) is given by \cite{kkbk,kik01}
\be
 {\hat D}_e({\hat x}) \equiv cst. e^{-\sum_{j=1}^N\frac{\omega^2}{2T}a_j^2-
\sum_{\alpha=1}^M\frac{1}{2T}b_{\alpha}^2}
\la{eqn:s4}
\ee
where $cst.$ is understood to be a suitably chosen constant. 

The mode-coupling coefficients $J_{ij\alpha}$ are
considered to be {\em static}  random variables satisfying the
following statistical properties:
\bea
\overline{J_{ij\alpha}}^J &=&0, \nonumber \\
\overline{J_{ij\alpha}J_{kl\beta}}^J &=& \frac{g^2}{N}
\biggl[(\delta_{ik}\delta_{jl}+\delta_{il}\delta_{jk})\delta_{\alpha
\beta}+K_{i\beta}(K_{k\al}\delta_{jl}+K_{l\al}\delta_{jk})
\nonumber \\ &+&
K_{j\beta}(K_{k\al}\delta_{il}+K_{l\al}\delta_{ik})\biggr]
\la{eqn:s5}
\eea
where $\overline{\cdots }^J$  is the average over the independent Gaussian
distribution of the $J$'s.
Eventually we take the mean field limit $M, N \rightarrow \infty$, keeping
$\delta^* \equiv M/N$ finite.

In writing the model equation the temperature and other model
parameters are fixed during time evolution, for example, in a situation
after the quench. Naturally the model equation is valid only in such a
time region.

\section{Action Integral}

In order to analyze the toy model in the limit $M, N \rightarrow \infty$,
 we introduce the following generating functional \cite{bckm}:
\bea
Z\{h^a,{\hat h}^a,h^b,{\hat h}^b \} &\equiv& \int d\{a\}\int
d\{b\}\int d\{\hat a\}\int d\{\hat b\}\nonumber \\
&\times& \exp \left\{i\int dt (h_j^aa_j+\hat
h_j^a\hat a_j +h_{\alpha}^bb_{\alpha}+ \hat h_{\alpha}^b\hat
b_{\alpha})\right\} \nonumber \\
&\times& e^{\hat{\cal S}_0+\hat{\cal S}_I} \la{eqn:s7}
\eea
where
\bea
\hat{\cal S}_0 &\equiv& \int dt\left\{ i\hat
a_i(\dot{a}_i-K_{i\alpha} b_{\alpha}) +i\hat
b_{\alpha}(\dot{b}_{\alpha} +\gamma b_{\alpha}+\omega^2K_{i\alpha
}a_i-f_{\alpha})\right\}(t) \nonumber \\
\hat{\cal S}_I&\equiv&J_{jk\alpha}X_{jk\alpha} \\
X_{jk\alpha}&\equiv &  \int dt \frac{\omega}{\sqrt{N}}\left\{-i
\hat a_j a_kb_{\alpha} + i\hat
b_{\alpha}(\omega^2a_ja_k-T\delta_{jk})\right\}(t)
 \la{eqn:s8}
\eea
In the above we have set the Jacobian of transformation of
variables to unity assuming the It\^{o} calculus$^{[1]}$. 
\footnotetext[1]{  A consequence of choosing the It\'o convention
is the causality condition on the response functions. 
This implies that when the responses of $a(t)$ or $b(t)$ to the disturbances
 $\hat a(t')$ or $\hat b(t')$
occur simultaneously, the limit $t'\rightarrow t$ must be chosen 
in such a way that $t$ is always {\it greater} than $t'$.}
In the limit $M,N \rightarrow \infty$, we find that the last term $-T\delta_{jk}$ 
in the integrand of (\ref{eqn:s8}) can be neglected, and will be dropped from now on.
That is, we take
\be
X_{jk\alpha}=  \int dt
\frac{\omega}{\sqrt{N}}\left\{-i \hat a_j a_kb_{\alpha}
+i \omega^2 \hat b_{\alpha}a_ja_k\right\}(t)
\la{eqn:s11}
\ee
We then notice that the replacements $ i\hat a_j \rightarrow
(\omega^2/T)a_j, \quad \hat b_{\alpha} \rightarrow b_{\alpha}/T$
on the rhs of (\ref{eqn:s11}) make this term vanish. Hence we can also
rewrite (\ref{eqn:s11}) as
\be
X_{jk\al}=  \int dt \frac{\omega}{\sqrt{N}}\left\{-i \tilde a_j a_kb_{\alpha}
+ i\omega^2 \tilde b_{\alpha}a_ja_k\right\}(t)
\la{eqn:s12}
\ee
where
\be
i\tilde a_j \equiv i\hat a_j+\frac{\omega^2}{T}a_j, \qquad i\tilde
b_{\alpha} \equiv i\hat b_{\alpha}+\frac{1}{T}b_{\alpha}.
\la{eqn:s13}
\ee
The quantities of interest are the out of equilibrium correlation
functions$^{[2]}$
\footnotetext[2]{The definitions of $C_{ab}$ and $C_{ba}$ and those of the $G$'s 
do not matter in the end. So we will use the symmetrically  defined ones.}
\bea
 C_a(t,t') &\equiv&  \frac{1}{N} <a_j(t)a_j(t')>, \quad
 C_{ab}(t,t') \equiv \frac{1}{M}K_{i\al}<a_i(t)b_{\al}(t')>,
 \nonumber \\
 C_{ba}(t,t') &\equiv& \frac{1}{M}K_{i\al}<b_{\al}(t)a_{i}(t')>,
 \quad
  C_b(t,t') \equiv \frac{1}{M} <b_{\alpha}(t)b_{\alpha}(t')> \nonumber
 \\
  C^K_a(tt') &\equiv& \frac{1}{M}<a^K_{\al}(t) a^K_{\al}(t')>, \quad
a^K_{\al} \equiv K_{j\al}a_j                                  
\la{eqn:s14}
\eea
and the response functions
\bea
 G_a(t,t') &\equiv& \frac{1}{N} <a_j(t)i{\hat a}_j(t')>, \quad
  G_{ab}(t,t') \equiv \frac{1}{M}K_{i\al}<a_i(t)i\hat
b_{\al}(t')>,  \nonumber \\
 G_{ba}(t,t') &\equiv& \frac{1}{M}K_{i\al}<b_{\al}(t)i\hat a_{i}(t')>,\quad
 G_b(t,t') \equiv \frac{1}{M} <b_{\alpha}(t)i {\hat b}_{\alpha}(t')>, \nonumber \\
 G^K_a(tt') &\equiv& \frac{1}{M}<a^K_{\al}(t)i\hat a^K_{\al}(t')>, \quad
 i\hat a^K_{\al} \equiv K_{j\al}i\hat a_j                                    
\la{eqn:s15}
\eea
We note that the new types of correlation and response functions $C^K_a$ and $G^K_a$  
are needed to obtain the closed set of equations for $C$'s and $G$'s for $M<N$.
For $M=N$ we have $C^K_a=C_a$ and $G^K_a=G_a$. 

Since we are here concerned with out-of-equilibrium situation we will not use
the time translation invariance nor 
the fluctuation-dissipation theorem (FDT)$^{[3]}$.
\footnotetext[3]{ The usual FDT takes the form where the
response function is proportional to the time derivative of the correlation function.
But here in the case of gaussian Hamiltonian \cite{dh75} the FDT is given by 
\begin{eqnarray*}
G_a(t-t')&=&-\theta(t-t')\frac{\omega^2}{T}C_a(t-t'), \quad
G_{ab}(t-t')=-\theta(t-t')\frac{1}{T}C_{ab}(t-t'), \nonumber \\
G_{ba}(t-t')&=&-\theta(t-t')\frac{\omega^2}{T}C_{ba}(t-t'), \quad
G_b(t-t')=-\theta(t-t')\frac{1}{T}C_{b}(t-t'), \nonumber \\
 G^K_a(t-t')&=&-\theta(t-t')\frac{\omega^2}{T}C^K_a(t-t') 
\end{eqnarray*}
where $\theta (t)$ is the usual step function equal to 1 for positive $t$ and zero otherwise,
the appearance of which comes from the causality. 
Another  property arising from the causality plus the above FDT is the following for arbitrary
$X(t)=X(a(t),b(t),\hat a(t), \hat b(t))$:
\[ <\hat A(t)X(t')>=<X(t) \tilde A(t')>=0 \quad \mbox{for} \quad t > t' \]
where $A(t)=(a(t),b(t))$ and the indices are suppressed for brevity. 
This fact is only limited to equilibrium when $h=\hat h=0$. 
 Hence  $ <X(t) \tilde A(t')>=0 \quad \mbox{for} \quad t > t'$ will  not be used here. 
For $\hat h=0$ and arbitrary $h$, however, the causality requires $ <\hat A(t)X(t')>=0 
\quad \mbox{for} \quad t > t'$, which will be used later. }
We now take averages of (\ref{eqn:s8}) over $f_{\alpha}$
and the $J$'s where we use
\begin{eqnarray}
\left<e^{-i \int dt \hat b_{\alpha}(t) f_{\alpha}(t)}\right>
&=&e^{-\gamma T \int dt \hat b_{\alpha}(t)^2} \nonumber \\
\overline{e^{J_{jk\alpha}X_{jk\alpha}}}^J&=&  e^{\frac{1}{2}
\overline{J_{jk\alpha}J_{lm\beta}}^J X_{jk\alpha}X_{lm\beta}}
 \la{eqn:s18}
\end{eqnarray}
Then we have
\be
<e^{\hat{\cal S}_0} > \equiv e^{{\cal S}_0}, \quad
 \overline{e^{\hat{\cal S}_I}}^J \equiv e^{{\cal S}_I}
\la{eqn:s19}
\ee
where
\bea
{\cal S}_0 &\equiv& \int dt\left\{ i\hat a_i(\dot{a}_i-K_{i\alpha}
b_{\alpha}) +i\hat b_{\alpha}(\dot{b}_{\alpha} +\gamma
b_{\alpha}+\omega^2K^T_{\alpha i}a_i+\gamma T i{\hat
b}_{\alpha})\right\}(t), \nonumber \\
 {\cal S}_I &\equiv&
\frac{g^2}{N} \int dt \int dt' \theta (t-t') \hat
\xi_{ij\alpha}(t) \biggl[ \bigl( \tilde \xi_{ij\alpha}(t')+ \tilde
\xi_{ji\alpha}(t') \bigr) \nonumber \\
 &+&  K_{i\beta}  K_{k\al} \bigl( \tilde
\xi_{kj\beta}(t')+\tilde \xi_{jk\beta}(t') \bigr) +
K_{j\beta} K_{k\al} \bigl( \tilde \xi_{ki\beta}(t')+
 \tilde \xi_{ik\beta}(t') \bigr)  \biggr]
\la{eqn:s20}
\eea
and
\begin{eqnarray}
 \hat \xi_{ij\al}(t) &\equiv& \hat \xi^a_{ij\al}(t)+\hat
 \xi^b_{ij\al}(t), \quad  \tilde \xi_{ij\al}(t') \equiv \tilde \xi^a_{ij\al}(t')
 +\tilde  \xi^b_{ij\al}(t') \nonumber \\
 \hat \xi^a_{ij\al}(t) &\equiv& \frac{\omega}{\sqrt{N}}(-i\hat
 a_ia_jb_{\al})(t), \quad \hat \xi^b_{ij\al}(t) \equiv \frac{\omega^3}{\sqrt{N}}
 (i\hat b_{\al}a_ia_j)(t) \nonumber \\
\tilde \xi^a_{ij\al}(t') &\equiv& \frac{\omega}{\sqrt{N}}(-i\tilde
 a_ia_jb_{\al})(t'), \quad \tilde \xi^b_{ij\al}(t') \equiv \frac{\omega^3}{\sqrt{N}}
 (i\tilde b_{\al}a_ia_j)(t').
 \la{eqn:s21}
\end{eqnarray}

Here we remind  that we can interchange
 $\hat \xi_{ij\alpha}$ and $\tilde \xi_{ij\alpha}$ in $\hat{\cal
 S}_I$, (\ref{eqn:s8}), as we have noted in connection with (\ref{eqn:s12}) and (\ref{eqn:s13}).
Therefore we can split ${\cal S}_I$ into the 4 contributions
\be
{\cal S}_I\equiv {\cal S}_I^{aa}+{\cal S}_I^{ab}+ {\cal S}_I^{ba}+{\cal S}_I^{bb}.
 \la{eqn:s22}
\ee
where
\begin{eqnarray}
&{}&{\cal S}_I^{aa} \equiv \frac{g^2}{N} \int dt \int dt' \theta
(t-t') \hat \xi^a_{ij\alpha}(t) \biggl[  \overbrace{\tilde
\xi^a_{ij\alpha}(t')}+ \tilde \xi^a_{ji\alpha}(t')
 \nonumber \\
 &+&  K_{i\beta} K_{k\al} \bigl( \tilde
\xi^a_{kj\beta}(t')+\tilde \xi^a_{jk\beta}(t') \bigr) +
K_{j\beta} K_{k\al} \bigl( \tilde \xi^a_{ki\beta}(t')+
  \overbrace {\tilde \xi^a_{ik\beta}(t')} \bigr) \biggr], \nonumber \\
&{}& {\cal S}_I^{ab} \equiv \frac{g^2}{N} \int dt \int dt' \theta
(t-t') \hat \xi^a_{ij\alpha}(t) \biggl[ \bigl( \tilde
\xi^b_{ij\alpha}(t')+ \tilde \xi^b_{ji\alpha}(t') \bigr)
 \nonumber \\
 &+&  K_{i\beta} K_{k\al} \bigl( \overbrace{\tilde
 \xi^b_{kj\beta}(t')}+\overbrace{\tilde \xi^b_{jk\beta}(t')} \bigr) +
 K_{j\beta}K_{k\al} \bigl( \tilde \xi^b_{ki\beta}(t')+
 \tilde \xi^b_{ik\beta}(t') \bigr) \biggr], \nonumber \\
&{}& {\cal S}_I^{ba} \equiv \frac{g^2}{N} \int dt \int dt' \theta
(t-t') \hat \xi^b_{ij\alpha}(t) \biggl[ \bigl( \tilde
\xi^a_{ij\alpha}(t')+ \tilde \xi^a_{ji\alpha}(t') \bigr) \nonumber \\
 &+& K_{i\beta} K_{k\al} \bigl( \overbrace{\tilde
\xi^a_{kj\beta}(t')}+\tilde \xi^a_{jk\beta}(t') \bigr) +
K_{j\beta} K_{k\al} \bigl( \overbrace{\tilde \xi^a_{ki\beta}(t')}+
 \tilde \xi^a_{ik\beta}(t') \bigr) \biggr], \nonumber \\
&{}&{\cal S}_I^{bb} \equiv \frac{g^2}{N} \int dt \int dt' \theta
(t-t') \hat \xi^b_{ij\alpha}(t) \biggl[ \bigl(\overbrace{ \tilde
\xi^b_{ij\alpha}(t')}+ \overbrace{\tilde \xi^b_{ji\alpha}(t')} \bigr) \nonumber \\
 &+&  K_{i\beta} K_{k\al} \bigl( \tilde
 \xi^b_{kj\beta}(t')+\tilde \xi^b_{jk\beta}(t') \bigr) +
 K_{j\beta} K_{k\al} \bigl( \tilde \xi^b_{ki\beta}(t')+
 \tilde \xi^b_{ik\beta}(t') \bigr) \biggr]
\la{eqn:s23}
\end{eqnarray}
Here the terms giving non-vanishing contributions in the equilibrium were overbraced.
But this is no longer enough in out of equilibrium situation as we shall
see. 
As an illustration we look at the first term of (\ref{eqn:s23}) or its integrand {${\cal S}_I^{aa}$}.
That  is
\bea 
\hat \xi^a_{ij\alpha}(t)\tilde \xi^a_{ij\alpha}(t')&=&\frac{\omega^2}{N}[-i\hat
 a_i(t)a_j(t)b_{\al}(t)][-i\tilde a_i(t')a_j(t')b_{\al}(t')]\nonumber \\
&=&\omega^2N \delta^* i\hat a_i(t)i\tilde
a_i(t')C_a(tt')C_b(tt')+\cdots
\la{eqn:s23a0}
 \eea
where the ellipsis contains quantities like $<i\hat a_i(t)\cdots>$
which are absent for $h,\,\,\hat h=0$ and also contains other
fluctuation terms. Due to the presence of a factor $N$ in front of the
second member of the rhs of (\ref{eqn:s23a0}), these fluctuation terms
disappear in the limit $M,N \rightarrow \infty$ with a finite $\delta^*$.
We then analyze each factor like $\hat \xi^a_{ij\alpha}(t)\tilde
\xi^a_{ij\alpha}(t')$   in (\ref{eqn:s23}) in  the limit $M,N
\rightarrow \infty$,
where we only impose the causality condition  $t \geq t'$ reflected by
$\theta(t-t')$ in (\ref{eqn:s23}) but not FDT.

After tedious but straightforward algebra with $h=0$ 
we arrive at the effective quadratic action given below. 
To simplify the expression we introduce the following notation$^{[4]}$:
$$ X \otimes Y(t) =\int_{-\infty}^t dt'X(tt')Y(t') $$
\footnotetext[4]{Note $\bigotimes$ signifies causality. Alternatively, this causality is
taken care of by redefining the $\Sigma$'s by absorbing $\theta(t-t')$
into them.  Then the above integral is from $-\infty$ to $\infty$. If
desired one can do the same for the response functions.}
In writing down the quadratic action below, for simplicity, 
we suppress time arguments and indices for the variables $a,b$,
so that for instance we write $\bf K\cdot b$ for $K_{j\al}b_{\al}$ in matrix notation.
We also omit integral signs and $\otimes$ for the moment.
The total action is given by the following matrix form
\bea
{\cal S}_{tot}&=&\hat{\cal S}_{eq}+{\cal S}_{eq}+{\cal S}_{oe}\nonumber\\
&\equiv&(i\hat{\bf a}, i\hat{\bf b})\cdot \hat \Omega_{eq}\cdot
\left(\begin{array}{rr} i\hat{\bf a} \\ i\hat{\bf b} \end{array}
\right)+(i\hat{\bf  a}, i\hat{\bf b})\cdot \Omega_{eq}\cdot
\left(\begin{array}{rr}{\bf a} \\ {\bf b} \end{array} \right)\nonumber
\\&+&(i\hat{\bf  a}, i\hat{\bf b})\cdot \Omega_{oe}^I\cdot
\left(\begin{array}{rr}{\bf a} \\ {\bf b} \end{array} \right)
\la{eqn:s32}
\eea
where [eq] and [oe] stand for equilibrium and off-equilibriun, respectively. 
Also note $\hat{\cal S}_{oe}$ is absent. 
We have
\bea
\hat{\Omega}_{eq}&\equiv&\hat{\Omega}_0+\hat{\Omega}_I\\
\hat{\Omega}_0 &\equiv&\left(\begin{array}{rr}{\bf 0}_N & {\bf 0}_{NM} \\ 
{\bf 0}_{MN} & \gamma T  {\bf 1}_M \end{array} \right) \\
\hat{\Omega}_I &\equiv&\left(\begin{array}{rr}{\bf 1}_N \frac{T}{\omega^2}
\Sigma_{aa}& T\Sigma_{ab}{\bf K} \\ 
\frac{ T}{\omega^2}\Sigma_{ba}{\bf K}^T&{\bf 1}_M
T\Sigma_{bb}\end{array} \right)
\la{eqn:bga1}
\eea
and furthermore,
\bea
\Omega_{eq}&=&\Omega_{eq}^0+\Omega_{eq}^I\\
{\Omega}_{eq}^0 &\equiv& \left(\begin{array}{rr}{\bf 1}_N\partial_t & -{\bf K}
 \\ \omega^2{\bf K}^T &    {\bf 1}_M (\partial_t+\gamma)\end{array}
\right)\\
\Omega_{eq}^I&\equiv&\left(\begin{array}{rr}{\bf 1}_N
\Sigma_{aa}& {\bf K} \Sigma_{ab} \\
 \Sigma_{ba}{\bf K}^T&{\bf 1}_M
\Sigma_{bb}\end{array} \right)
\la{eqn:bga2}
\eea
and
\be
\Omega_{oe}^I\equiv\left(\begin{array}{rr}{\bf 1}_N
(-\Sigma_{aa}+\Delta  \Sigma_{aa}-2\Delta \Sigma_{aa}^{\odot})& 
-( \Sigma_{ab}+\Delta  \Sigma_{ab}){\bf K} \\
(-\Sigma_{ba}+\Delta \Sigma_{ba}-2\Delta \Sigma_{ba}^{\odot}){\bf K}^T&-{\bf 1}_M
(\Sigma_{bb}+\Delta  \Sigma_{bb})\end{array} \right)
\la{eqn:bga3}
\ee
where ${\bf 1}_N({\bf 1}_M)$ are  the unit matrix of rank $N(M)$, $\bf K$ is 
the $N \times M$ matrix with the elements $K_{j\al}$ and ${\bf K}^T$ its transposed $M \times N$ matrix.
All these matrices are multiplied by a matrix whose $tt'$ element is the delta function
$\delta (t-t')$.
Also ${\bf 0}_{NM},\,{\bf 0}_{MN}$ are the 0 matrices of $N \times M,
\,M \times N$, respectively.
Here the memory kernel $\Sigma$'s  are  defined by 
\bea
\Sigma_{aa}(tt') &\equiv&
\frac{g^2\omega^4}{T}\bigl(\delta^*C_a(tt')C_b(tt')+(\delta^*)^2C_{ab}(tt')C_{ba}(tt')\bigr)\nonumber\\ 
 \Sigma_{ab}(tt') &\equiv& -2\frac{g^2\omega^4}{T}\delta^*C_a(tt')C_{ba}(tt'), \quad
\Sigma_{ba}(tt') \equiv -2 \frac{g^2\omega^6}{T}\delta^* C_a(tt')C_{ab}(tt'), \nonumber \\
 \Sigma_{bb}(tt')&\equiv& \frac{2g^2\omega^6}{T}C_{a}(tt')^2
\la{eqn:s31}
\eea
The other types of kernels $\Delta \Sigma$ and  $\Delta \Sigma^{\odot}$
are defined as follows:
\bea
\Delta \Sigma_{aa}&\equiv&{g^2\omega^2}\bigl(\delta^*
G_aC_b+(\delta^*)^2C_{ab}G_{ba}\bigr)
\la{eqn:bga3aa}\\
\Delta \Sigma^{\odot}_{aa}&\equiv&{g^2\omega^4}\bigl(\delta^*
C_a G_b + (\delta^*)^2 C_{ba}G_{ab}\bigr)
\la{eqn:bga3ba}\\
\Delta \Sigma_{ab}&\equiv&-g^2\omega^2\delta^*[C_aG_{ba}+G_aC_{ba}]
\la{eqn:bga3ab}\\
\Delta \Sigma_{bb}&\equiv& 2g^2\omega^4 C_aG_a, \quad 
\Delta \Sigma_{ba}\equiv-2g^2\omega^4  \delta^* 
C_{ab}G_{a}
\la{eqn:bga3ac}\\
\Delta \Sigma^{\odot}_{ba}&\equiv&-2g^2\omega^6\delta^*C_aG_{ab}
\la{eqn:bga3ad}
\eea
Note that every term of the $\Delta \Sigma$'s contains one factor of
the $G$'s and one factor of $C$'s in contrast to the $\Sigma$'s, (\ref{eqn:s31}), each of which contains the
two factors of $C$'s. 
\section{CORRELATION AND RESPONSE FUNCTIONS}
We now proceed to response and correlation functions.
 We first introduce correlation and response matrices $\bf C$ and
 $\bf G$ with sub-matrices ${\bf C}_{aa}\equiv <\bf aa>$ and ${\bf
 G}_{aa}\equiv <{\bf a}i\hat{\bf a}>$ etc whose elements
 are $<{\bf aa}>_{it,jt'}=<a_i(t)a_j(t')>$ etc.
 Thus the entire correlation and response matrices are written as
\be
{\bf C}\equiv  \left(\begin{array}{rr}{\bf C}_{aa} & {\bf C}_{ab} \\
 {\bf C}_{ba} &{\bf C}_{bb} \end{array} \right),\quad {\bf G}\equiv
  \left(\begin{array}{rr}{\bf G}_{aa} &{\bf G}_{ab}  \\
{\bf G}_{ba} & {\bf G}_{bb} \end{array} \right)
\la{eqn:aga21}
\ee
The formal matrix equations determining correlation and response
 matrices take the form which are obtained from the effective action
 defined through (\ref{eqn:s32}) to (\ref{eqn:bga3}):
\bea
\Omega \cdot {\bf G}&=&{\bf 1}
\la{eqn:aga22a}\\
\Omega \cdot {\bf C}&=&\big(\hat \Omega_{eq}+\hat
 \Omega_{eq}^{\dagger}\big)\cdot {\bf G}^{\dagger}
\la{eqn:aga22b}\\
\Omega &\equiv& \Omega_{eq}+\Omega_{oe}^I
\la{eqn:aga22c}
\eea
\subsection{Response Function}
We first take up the response functions. 
The equations for them are written in terms of submatrices as follows where
 equilibrium and off-equilibrium parts are separated:
\bea
&{}&(\partial_t+ \Sigma_{aa}){\bf G}_{aa}-(1- \Sigma_{ab}){\bf
  K}\cdot{\bf G}_{ba} \nonumber\\& +&[[(- \Sigma_{aa}+\Delta \Sigma_{aa}-2\Delta
\Sigma_{aa}^{\odot}){\bf G}_{aa}-(\Sigma_{ab}+\Delta \Sigma_{ab}){\bf
  K\cdot G}_{ba})]]\nonumber \\&=&{\bf 1}_N\la{eqn:aga24aa}\\
&{}&(\partial_t+ \Sigma_{aa}){\bf G}_{ab}-(1-\Sigma_{ab}){\bf
  K}\cdot{\bf G}_{bb} \nonumber \\&+&[[(- \Sigma_{aa}+\Delta \Sigma_{aa}-2\Delta
\Sigma_{aa}^{\odot}){\bf G}_{ab}-(\Sigma_{ab}+\Delta \Sigma_{ab}){\bf
  K\cdot G}_{bb})]]\nonumber\\&=&{\bf 0}_{NM}\la{eqn:aga24ab}\\
&{}&(\omega^2+ \Sigma_{ba}){\bf K}^T\cdot{\bf
  G}_{aa}+(\partial_t+\gamma+\Sigma_{bb}){\bf G}_{ba}\nonumber\\&+&[[(-
  \Sigma_{ba}+\Delta \Sigma_{ba}-2\Delta \Sigma_{ba}^{\odot}){\bf
  K}^T\cdot {\bf G}_{aa}-(\Sigma_{bb}+\Delta \Sigma_{bb})\cdot {\bf
  G}_{ba}]]\nonumber\\ &=& {\bf 0}_{MN}\la{eqn:aga24ac}\\
&{}&(\omega^2+ \Sigma_{ba}){\bf K}^T\cdot{\bf
  G}_{ab}+(\partial_t+\gamma+\Sigma_{bb}){\bf G}_{bb}\nonumber\\&+&[[(-
  \Sigma_{ba}+\Delta \Sigma_{ba}-2\Delta \Sigma_{ba}^{\odot}){\bf
  K}^T\cdot {\bf G}_{ab}-(\Sigma_{bb}+\Delta \Sigma_{bb})\cdot {\bf
  G}_{bb}]]\nonumber\\ &=& {\bf 1}_{M}
\la{eqn:aga24ad}
\eea
Here $[[\cdots]]$ are off-equilibrium parts which vanish in equilibrium
due to the FDT which makes all the $\Delta \Sigma$'s and 
$\Delta \Sigma^{\odot}$ equal to $-\Sigma$'s.
From this we can deduce the equations for 5 response functions in the
following manner.

We define the following notation valid for arbitrary matrix $\bf X,Y$
 of ranks $N,M$, respectively:
$$\tr^a {\bf X}\equiv \frac{1}{N}\sum_j X_{jj}, \quad 
\tr^b {\bf Y}\equiv \frac{1}{M}\sum_{\al} Y_{\al\al}$$
We first apply $\tr^a \cdots$ to (\ref{eqn:aga24aa}), 
next $\tr^b{\bf K}^T\cdots$ to (\ref{eqn:aga24ab}),
then  $\tr^a {\bf K} \cdots$ to (\ref{eqn:aga24ac}), also
$\tr^b\cdots$ to (\ref{eqn:aga24ad}), and
finally $\tr^b {\bf K}^T\cdots {\bf K}$ to (\ref{eqn:aga24aa}). 

We then end up with  the following set of 5 equations for 5 response functions
where again  equilibrium and off-equilibrium parts are separated:

\smallskip
\bea
&{}&[ (\partial_t + \Sigma_{aa})G_a
-(1- \Sigma_{ab})\delta^* G_{ba}](tt')\nonumber\\&+&[[(-\Sigma_{aa}+\Delta
\Sigma_{aa}-2\Delta \Sigma_{aa}^{\odot})G_a-(\Sigma_{ab}+\Delta
\Sigma_{ab})\delta^*G_{ba}]](tt')\nonumber \\&=&\delta(t-t')
\la{eqn:aga26aa}\\
&{}& [(\partial_t + \Sigma_{aa})G_{ab}
-(1- \Sigma_{ab}) G_{b}](tt')\nonumber\\&+&[[(-\Sigma_{aa}+\Delta
\Sigma_{aa}-2\Delta \Sigma_{aa}^{\odot})G_{ab}-(\Sigma_{ab}+\Delta
\Sigma_{ab})G_{b}]](tt')\nonumber\\&=&0 
 \la{eqn:aga26ab}\\
&{}& [(\omega^2+\Sigma_{ba})G_a^K+(\partial_t+\gamma+\Sigma_{bb})
G_{ba}](tt')\nonumber\\&+&[[(-\Sigma_{ba}+\Delta\Sigma_{ba} -2\Delta
\Sigma_{ba}^{\odot}) G_{a}^K-(\Sigma_{bb}+\Delta \Sigma_{bb})G_{ba}
]](tt')\nonumber \\&=&0 
\la{eqn:aga26ac}\\
&{}&[ (\omega^2+\Sigma_{ba})G_{ab}+(\partial_t+\gamma+\Sigma_{bb})
G_b](tt')\nonumber\\&+&[[(-\Sigma_{ba}+\Delta\Sigma_{ba} -2\Delta
\Sigma_{ba}^{\odot}) G_{ab}-(\Sigma_{bb}+\Delta \Sigma_{bb})G_{b}
]](tt')\nonumber\\&=&\delta(t-t')
 \la{eqn:aga26ad}\\
&{}&[(\partial_t+\Sigma_{aa})G_{a}^K-(1- \Sigma_{ab})G_{ba}](tt')\nonumber\\
&+&[[(-\Sigma_{aa}+\Delta\Sigma_{aa} -2\Delta \Sigma_{aa}^{\odot})G_a^K
-(\Sigma_{ab}+\Delta \Sigma_{ab})G_{ba}]](tt')\nonumber\\&=&\delta(t-t')
\la{eqn:aga26ae}\\ \nonumber
\eea

\subsection{Correlation Functions}
We start with the formula (\ref{eqn:aga22b})
where the lhs are the same as those of (\ref{eqn:aga24aa}) to (\ref{eqn:aga24ad})
except that the $\bf G$'s are replaced by the $\bf C$'s. 
Thus we need to consider only the rhs.
Note that the rhs is the same as the equilibrium case if  ${\bf G}^{\dagger}(tt')$ is given.
${\bf G}^{\dagger}(tt')$ in terms of submatrices is given by
\be
{\bf G}^{\dagger}(tt') =
\left(\begin{array}{rr}<i\hat {\bf a}(t'){\bf a}(t)> & <i\hat {\bf a}(t'){\bf b}(t)> \\
<i\hat{\bf b}(t'){\bf a}(t)> & <i\hat{\bf b}(t') {\bf b}(t)> \end{array} \right)
\la{eqn:cor2}
\ee
Also one can work out
\bea
(\hat\Omega_{eq}+\hat\Omega_{eq}^{\dagger})(tt')
&=& 
\left(\begin{array}{rr}\frac{T}{\omega^2}
     [\Sigma_{aa}(tt')+\Sigma_{aa}(t't)]{\bf 1}_N, &
 T [\Sigma_{ab}(tt')+\frac{1}{\omega^2}\Sigma_{ba}(t't)]{\bf K}\\
 T [\frac{1}{\omega^2}\Sigma_{ba}(tt')+\Sigma_{ab}(t't)]{\bf K}^T, & 
T [2\gamma \delta(t-t')+\Sigma_{bb}(tt')+\Sigma_{bb}(t't)]{\bf 1}_M
\end{array} \right) \nonumber \\
&=& 
\left(\begin{array}{rr}2\frac{T}{\omega^2}
     \Sigma_{aa}(tt'){\bf 1}_N, &
 2T \Sigma_{ab}(tt'){\bf K}\\
 2T \Sigma_{ab}(t't){\bf K}^T, & 
2T [\gamma \delta(t-t')+\Sigma_{bb}(tt')]{\bf 1}_M
\end{array} \right) 
\la{eqn:cor2a}
\eea
The last equality in (\ref{eqn:cor2a}) is due to 
the  following symmetric properties of $\Sigma$'s under exchange of two times:
$\Sigma_{aa}(t,t')=\Sigma_{aa}(t',t) $, $\Sigma_{ab}(t,t')=\frac{1}{\omega^2} \Sigma_{ba}(t',t)$, and  
$\Sigma_{bb}(t,t')=\Sigma_{bb}(t,'t) $.
These follow directly from $C_a(t,t)=C_a(t',t)$,  $C_b(t,t)=C_b(t',t)$, and $C_{ab}(t,t')=C_{ba}(t',t)$.
The matrix equation (\ref{eqn:aga22b}) can be
split into 4 submatrix equations as follows  where the lhs is
abbrevitated as $\partial_t{\bf C}_{aa}(tt') +\cdots$ etc.

\bea
\partial_t{\bf C}_{aa}(tt') +\cdots\ &=& 2 \frac{T}{\omega^2}
\Sigma_{aa}(t \bullet)<i\hat {\bf a}(\bullet){\bf a}(t')>\nonumber \\
&+& 2 T \Sigma_{ab}(t \bullet){\bf K}\cdot<i\hat{\bf b}(\bullet){\bf a}(t')>,
\la{eqn:aga28a}\\
\partial_t{\bf C}_{ab}(tt') +\cdots&=&2\frac{T}{\omega^2}
\Sigma_{aa}(t \bullet) <i\hat {\bf a}(\bullet){\bf b}(t')>\nonumber \\
&+&2 T \Sigma_{ab}(t \bullet){\bf K}\cdot<i\hat{\bf b}(\bullet){\bf b}(t')>,
\la{eqn:aga28b}\\
\partial_t{\bf C}_{ba}(tt') +\cdots &=&2 T
\Sigma_{ab}(\bullet t){\bf K}^T\cdot <i\hat {\bf a}(\bullet){\bf a}(t')>\nonumber\\
&+&2T [\gamma \delta (t-\bullet)+\Sigma_{bb}(t \bullet)] <i\hat{\bf b}(\bullet){\bf a}(t')>, 
\la{eqn:aga28c} \\
\partial_t{\bf C}_{bb}(tt') +\cdots&=&2T \Sigma_{ab}(\bullet t){\bf K}^T\cdot
 <i\hat {\bf a}(\bullet){\bf b}(t')> \nonumber \\
&+&2T [\gamma \delta (t-\bullet)+\Sigma_{bb}(t \bullet)] <i\hat{\bf b}(\bullet){\bf b}(t')>
\la{eqn:aga28d} \\ \nonumber 
\eea
The rhs are the same with the equilibrium case.

We are ready to find the rhs of the equations for correlation functions,
which can be done following the same procedure as for the $G$'s.
Thus we first apply $\tr^a \cdots$ to (\ref{eqn:aga28a}), 
next $\tr^b{\bf K}^T\cdots$ to the rhs of (\ref{eqn:aga28b}),
then $\tr^a{\bf K}\cdots$ to the rhs of (\ref{eqn:aga28c}), 
also $\tr^b\cdots$ to the rhs of (\ref{eqn:aga28d}), and
finally $\tr^b {\bf K}^T\cdots {\bf K}$ to the rhs of
(\ref{eqn:aga28a}).

Results are summarized below in the form of 5 self-consistent equations for 5 correlation functions:
\smallskip
\bea 
&{}&[(\partial_t + \Sigma_{aa})C_a -(1- \Sigma_{ab})\delta^* C_{ba}](tt')  \nonumber\\
&+&[[(-\Sigma_{aa}+\Delta \Sigma_{aa}
-2\Delta \Sigma_{aa}^{\odot})C_a-(\Sigma_{ab}+\Delta \Sigma_{ab})\delta^*C_{ba}]](tt')\nonumber\\
&=&2\frac{T}{\omega^2} \Sigma_{aa}(t\bullet)G_a(t'\bullet)
+2T\Sigma_{ab}(t\bullet)\delta^*G_{ab}(t'\bullet),
\la{eqn:aga30a} \\ 
&{}& [(\partial_t + \Sigma_{aa})C_{ab}
-(1- \Sigma_{ab}) C_{b}](tt')\nonumber\\
&+&[[(-\Sigma_{aa}+\Delta \Sigma_{aa}-2\Delta \Sigma_{aa}^{\odot})C_{ab}-
(\Sigma_{ab}+\Delta \Sigma_{ab})C_{b}]](tt')\nonumber\\  
&=&2\frac{T}{\omega^2} \Sigma_{aa}(t\bullet)G_{ba}(t'\bullet)+
2T \Sigma_{ab}(t\bullet)G_{b}(t'\bullet), 
\la{eqn:aga30b} \\
&{}& [(\omega^2+\Sigma_{ba}) C_a^K+(\partial_t+\gamma+\Sigma_{bb})
C_{ba}](tt')\nonumber\\&+&[[(-\Sigma_{ba}+\Delta\Sigma_{ba} -2\Delta
\Sigma_{ba}^{\odot}) C_{a}^K-(\Sigma_{bb}+\Delta \Sigma_{bb})C_{ba}]](tt')\nonumber \\
&=&2T\Sigma_{ab}(\bullet t) G_a^K(t'\bullet) 
+2T [\gamma \delta (t-\bullet)+\Sigma_{bb}(t\bullet)] G_{ab}(t'\bullet), 
\la{eqn:aga30c} \\
&{}&[ (\omega^2+\Sigma_{ba})C_{ab}+(\partial_t+\gamma+\Sigma_{bb}) C_b](tt')\nonumber\\
&+&[[(-\Sigma_{ba}+\Delta\Sigma_{ba} -2\Delta \Sigma_{ba}^{\odot})C_{ab}
-(\Sigma_{bb}+\Delta \Sigma_{bb})C_b]](tt')\nonumber\\
&=&2T \Sigma_{ab}(\bullet t)]G_{ba}(t'\bullet)
+2T [\gamma \delta (t-\bullet)+\Sigma_{bb}(t\bullet)]G_b(t'\bullet), 
\la{eqn:aga30d} \\ 
&{}&[(\partial_t+\Sigma_{aa})C_{a}^K-(1- \Sigma_{ab})C_{ba}](tt')\nonumber\\
&+&[[(-\Sigma_{aa}+\Delta\Sigma_{aa} -2\Delta \Sigma_{aa}^{\odot})C_a^K
-(\Sigma_{ab}+\Delta \Sigma_{ab})C_{ba}]](tt')\nonumber\\ 
&=&2\frac{T}{\omega^2} \Sigma_{aa}(t\bullet)G_a^K(t'\bullet) 
+2T\Sigma_{ab}(t\bullet)G_{ab}(t'\bullet) 
\la{eqn:aga30e} 
\eea
\smallskip

The 5 equations (\ref{eqn:aga26aa})-(\ref{eqn:aga26ae}) and 5 equations
(\ref{eqn:aga30a})-(\ref{eqn:aga30e})  constitute 10 equations that
self-consistently determine 5 correlation functions and 5 response
functions. Note the rhs of (\ref{eqn:aga30a})-(\ref{eqn:aga30e})
are the same as in equilibrium case. 
The lhs again have been split up into equilibrium and nonequilibrium portions
as we have done in (\ref{eqn:aga26aa})-(\ref{eqn:aga26ae}).

\noindent
We now give several comments related to the set of equations (\ref{eqn:aga30a})-(\ref{eqn:aga30e}):
\begin{itemize}
\item So far we have considered a general out of equilibrium situation where
the model parameter remains constant. Now we specify the condition to
meet a typical aging experiment. We then suppose that a system in
some equilibrium state is quenched at the time $t=0$ and starts to
evolve into another equilibrium state which characterizes the model
parameters. We let the system age till some time $t_w(>0)$, and we
measure at a later time $t(>t_w)$. This would mean that in the above
equations we should take $t'=t_w$ and the time integrals denoted by
$\bullet \equiv s$ should be over the region $s>0$, which is further limited
by the causality conditions on the $\Sigma$'s and the $G$'s.
Thus each integral in the equations of the response functions, 
(\ref{eqn:aga26aa})-(\ref{eqn:aga26ae}),  is in the interval $t_w <s < t$. 
Similarly, each integral in the lhs of the equations for the correlation
functions, (\ref{eqn:aga30a})-(\ref{eqn:aga30e}), is in the interval
$0<s<t$ whereas each integral in the rhs is in the interval 
$0<s<t_w$. 
We then observe that the rhs of  (\ref{eqn:aga30a}) to (\ref{eqn:aga30e}) are
the source of contributions to the correlation functions from thermal noise
generated after the quench. Note that the functions that multiply the
$G(t'\bullet)$'s in  the rhs of  (\ref{eqn:aga30a}) to
(\ref{eqn:aga30e}) are the correlation functions of renormalized
thermal noises. See the equations (25) of \cite{kik01}.
\item  
In the discussion above we have not considered effects of the initial
condition at the time of quench, say, $t_0$ \cite{tk88}. This can be
studied by inserting the properly normalized weight factor
proportional to $\exp(-H/k_BT_0)$ where $H$ is the system
Hamiltonian and $T_0$ is the initial temperature. For spin glass cases including
Potts or $p$-spin systems, $H$ contains quenched random interaction
parameters, which have to be included in averaging the exponential of
the action integral over such quenched random parameters contained in
the action integral. This results in additional terms in the final
effective action. In our toy model, this complication is absent
because the Hamiltonian is free from quenched disorder. 
The initial condition enters only
through the initial values of the correlation functions, that is,
$C_a(t=0,t_w=0)=T_0/\omega^2$, etc.
\item  
If we take all the $\Delta \Sigma$'s and $\Delta
\Sigma^{\odot}$equal to $-\Sigma$'s, consequences of FDT, we recover the equilibrium
equations. In this case the rhs of 5 equations  (\ref{eqn:aga30a})-(\ref{eqn:aga30e})
 must vanish for $t>t'$ because of the FDT
which tells that the $G$'s to be proportional to the $C$'s. The direct
verification of this follows if we note that in equilibrium $t_w$ can
be shifted to $0$. This makes the integration interval of $s$, and
hence the rhs of  (\ref{eqn:aga30a}) to (\ref{eqn:aga30e}) to vanish. 
\item 
 No $\Delta \Sigma_{ab}^{\odot}$ appears in constrast to 
$\Delta \Sigma_{ba}^{\odot}$, which is due to asymmetric way $a$ and $b$ variables enter dynamics.
\item   No $C_a^K$ and  $G_a^K$ appear in the $\Sigma$'s.
One can see this by inspecting the derivations of (\ref{eqn:s32}) to (\ref{eqn:bga3}). 
We see that no terms containing the combinations\\ $a^K_{\al}(t)a^K_{\al}(t'),\quad
a^K_{\al}(t)i\hat a^K_{\al}(t')$ with $a^K_{\al} \equiv K_{j\al}a_j,
\quad  i\hat a^K_{\al} \equiv K_{j\al}i\hat a_j $ appear.
\end{itemize}
 
\section{Concluding remarks}
In the foregoing sections we have derived the exact self-consistent
equations for correlation functions and response functions for our
toy model which, in some sense, is complementary to the works of
Latz \cite{latz} who studied the microscopic fluid system.

The set of ten equations of our self-consistent scheme would be too
complicated for further analyses. Now, the velocity-like $b$-variables
can be made rapidly decaying by choosing a sufficiently large value
for $\gamma$. In particular, this gives rise to a possibility of
adiabatically eliminating the velocity-like variables as we have done
to derive a Fokker-Planck type equation for the probability
distribution function of $a$-variables only in \cite{kik01,merida}.         

\medskip
\begin{acknowledgments}
The research of KK is supported by the Department of Energy, under contract
W-7405-ENG-36. An additional partial support to KK by the Cooperative
Research under the Japan-U.S. Cooperative Science Program sponsored by Japan Society
of Promotion of Science is  also gratefully acknowledged.
BK is supported by the Interdisciplinary Research Program of the KOSEF (Grant No.
1999-2-114-007-3).
\end{acknowledgments}

\end{document}